\documentclass[conference]{IEEEtran}
\IEEEoverridecommandlockouts

\usepackage{cite}
\usepackage{amsmath,amssymb,amsfonts}
\usepackage{algorithmic}
\usepackage{graphicx}
\usepackage{textcomp}
\usepackage{xcolor}
\usepackage[linesnumbered,ruled]{algorithm2e}
\usepackage{subfigure}
\usepackage{caption}
\usepackage{balance}

\newcommand{\eg}{\emph{e.g.}\xspace}
\newcommand{\ie}{\emph{i.e.}\xspace}

\def\BibTeX{{\rm B\kern-.05em{\sc i\kern-.025em b}\kern-.08em
    T\kern-.1667em\lower.7ex\hbox{E}\kern-.125emX}}

\begin{document}

\setlength{\textfloatsep}{8pt plus 1.0pt minus 2.0pt}

\title{Synergistic CPU-FPGA Acceleration of \\ Sparse Linear Algebra}

\author{\IEEEauthorblockN{Mohammadreza Soltaniyeh}
\IEEEauthorblockA{\textit{Department of Computer Science} \\
\textit{Rutgers University}\\
Piscataway, USA \\
m.soltaniyeh@cs.rutgers.edu}
\and
\IEEEauthorblockN{Richard P. Martin}
\IEEEauthorblockA{\textit{Department of Computer Science} \\
\textit{Rutgers University}\\
Piscataway, USA \\
rmartin@scarletmail.rutgers.edu
}
\and
\IEEEauthorblockN{Santosh Nagarakatte}
\IEEEauthorblockA{\textit{Department of Computer Science} \\
\textit{Rutgers University}\\
Piscataway, USA \\
santosh.nagarakatte@cs.rutgers.edu \vspace{0.2cm}}

\hspace{-4.5 in} {\large \bf Rutgers Computer Science Technical Report DCS-TR-750} }

\IEEEpeerreviewmaketitle
\maketitle

\begin{abstract}

This paper describes REAP, a software-hardware approach that enables
high performance sparse linear algebra computations on a cooperative
CPU-FPGA platform.  REAP carefully separates the task of organizing
the matrix elements from the computation phase.  It uses the CPU to
provide a first-pass re-organization of the matrix elements, allowing
the FPGA to focus on the computation. We introduce a new intermediate
representation that allows the CPU to communicate the sparse data and
the scheduling decisions to the FPGA. The computation is optimized on
the FPGA for effective resource utilization with pipelining.  REAP
improves the performance of Sparse General Matrix Multiplication
(SpGEMM) and Sparse Cholesky Factorization by 3.2$\times$ and
1.85$\times$ compared to widely used sparse libraries for them on the
CPU, respectively.

\end{abstract}

\section{Introduction}

Sparse linear algebra operations are important in many domains,
including simulating physical body dynamics~\cite{hussein2011sparse},
multigrid methods~\cite{georgii2010streaming}, network
routing\cite{zwick2002all}, and integer
factorization~\cite{bouvier2013filtering}. The difference in delivered
Giga-Floating Point Operations (GFLOPs) between dense and sparse codes
is 10-50$\times$ depending on the sparsity pattern.  An exacerbating
issue impacting performance is the need to keep the matrices in the
sparse formats that involve a level of indirection.  The challenge is
both finding and matching matrix elements which need to be operated
on, which is an extra step compared to the dense case. Indeed, the
overhead of accessing the indices into the sparse structure, not
including the matching cost, can exceed the work of the mathematical
operations by factors of 2-5$\times$~\cite{im2004sparsity}.

Our work is motivated by the promise of Field Programmable Gate Arrays
(FPGAs) as computation engines for high-performance computing. With
thousands of functional units and their ability to internally route
data in an application-dependent manner, FPGAs present an opportunity
to scale performance and obtain better performance-per-watt. An
important challenge in accelerating sparse computation with FPGAs is
the need to keep the computation units busy. A level of indirection
with sparse computation hinders performance with an FPGA.

This paper introduces REAP, a co-operative CPU-FPGA approach for
sparse computation that takes effectively uses the strengths of both
the CPU and FPGA. With many layers of caches, CPUs are good at
manipulating small-scale, unpredictable memory access patterns.
In contrast, FPGA can be synthesized with a large number of functional
units and can be re-programmed to have application-specific routing
and logic. REAP uses the CPU to re-organize the sparse matrices into a
form that can be streamed into the FPGA. We also show how the CPU
providing the FPGA with metadata about the element location aids it in
matching operands.

REAP takes as input matrices in standard formats, such as Compressed
Sparse Row (CSR) and Coordinate Format (COO). Using a standard format
enhances portability, maintenance, and data curation. The CPU
interfaces with the FPGA using our proposed new intermediate
representation, which we call REAP Intermediate Representation~(RIR).
CPU rearranges the sparse matrix elements into discrete chunks, which
we call RIR bundles.  A bundle contains metadata and the matrix
elements that share a feature, such as a row or a column.
The FPGA can then route the RIR bundles to pipelines of processing
elements (PE) which process the data in the bundle to create partial
results. The partial results are also maintained in bundles, which can
then be sent to additional units to be matched to results from
successive bundles. We have generalized the RIR bundles for both
SpGEMM and sparse Cholesky factorization.
This format allows a transformation of a sparse pattern into a linear
sequence of bundles. The linearization of the matrix elements means
the FPGA can stream the elements from the main memory to obtain a very
effective high bandwidth. For Cholesky factorization, we found that
adding some metadata-only bundles further improves performance.

Using a combination of synthesis from Hardware Description Languages
(HDLs) and simulation, our results show that the co-operative approach
of using a CPU to precondition the matrix elements followed by
streaming them into the FPGA outperforms a multi-core CPU even at
modest matrix element densities. In particular, we show we can always
obtain speedups over CPUs when the density ratio is over 1:1000 for
both Sparse Generalized Matrix Multiplication~(SpGEMM) and Cholesky
factorization.  We observe a geometric mean speedup of 3.2$\times$
over a single CPU version using Intel's Math Kernel Library for
SpGEMM.
We observed that our designs attain a geometric mean speedup of
1.8$\times$ with Cholesky factorization over a single CPU that uses
the CHOLMOD library~\cite{CHOLMOD:2008}.

A surprising finding is that REAP is faster than a multi-core CPU with
an equal number of floating point units in the FPGA and CPU.  Our
hypothesis for this behavior is that streaming reformatted bundles
into the FPGA results in better throughput through the floating point
units compared to moving data through the CPU's caches.
An important finding of our work is these speedups are not obtainable
without sufficient bandwidth between the memory and FPGA

\textbf{Summary.} REAP's co-operative CPU-FPGA approach increases the
effective bandwidth of the FPGA design by letting the CPU performs the
scheduling and marshaling of the input. Hence, FPGA enjoys regular
accesses to memory. Further, relegating all the computation to the
FPGA where there are abundant computation resources including
floating-point DSP units improves performance. Finally, decoupling the
CPU and FPGA's work in a coarse-grained fashion allows their
overlapped execution for higher performance.

\section{Synergistic CPU-FPGA Acceleration}

\begin{figure}[t]
  \begin{center}
    
  \includegraphics[width=0.5\textwidth,angle=0]{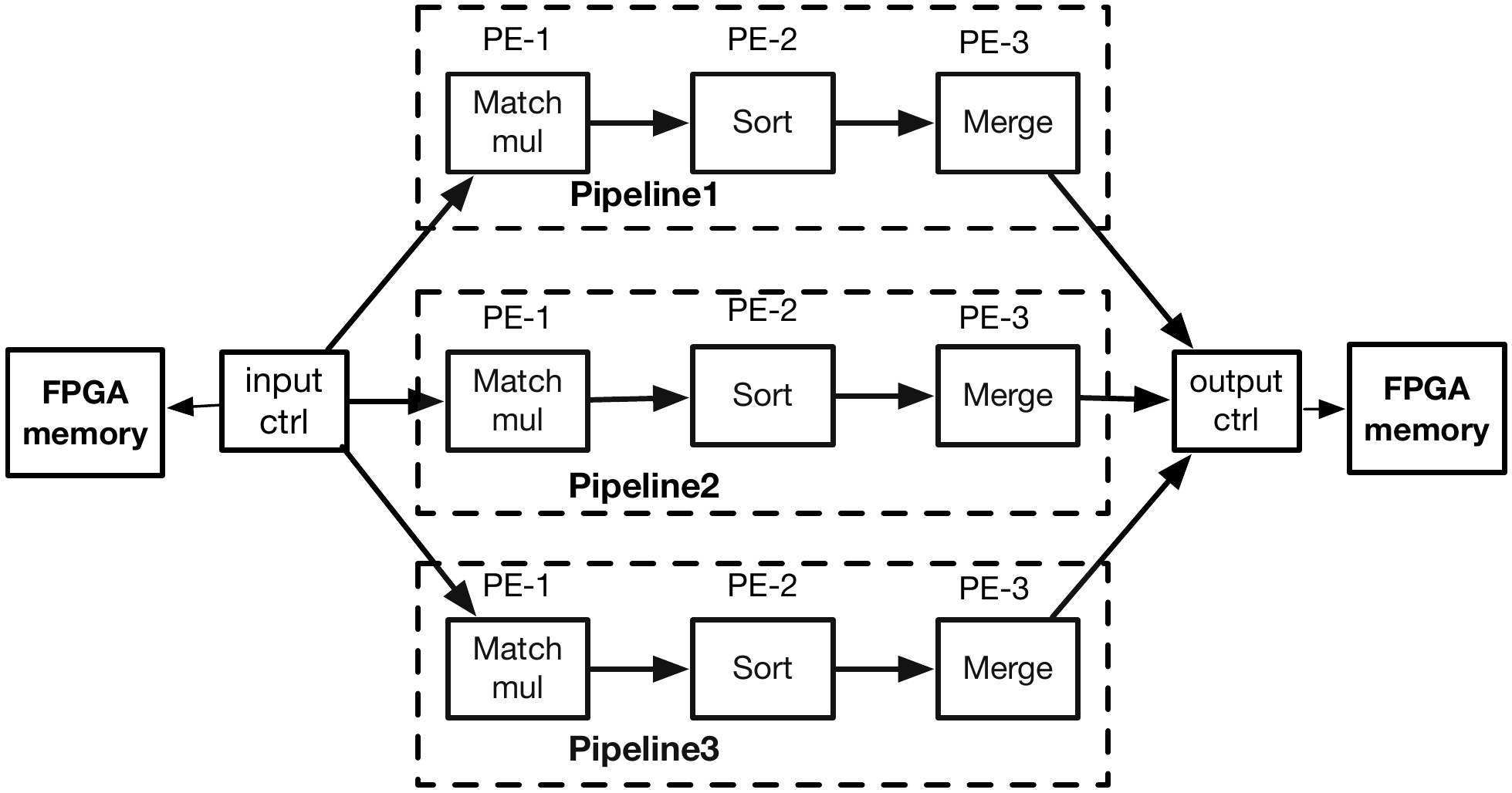}
  \caption{REAP FPGA Design for SpGEMM}
  \label{fig:spgemmarchdesign}

\end{center}
\end{figure}

This section provides an overview of REAP. We first show the overall
architecture of the design and then describe how sparse data is
regularized for the FPGA.

\textbf{High-Level FPGA design in REAP.}

Figure~\ref{fig:spgemmarchdesign} provides an overview of the
organization of the computation in the FPGA for SpGEMM. The Cholesky
factorization is similar. It is described in detail in
Section~\ref{sec:cholesky}.  Both designs center around replicated
pipelines, which are organized vertically in
Figure~\ref{fig:spgemmarchdesign}. Pipelines consist of multiple
stages; there is thus parallelism with regards to the number of
pipelines we can fit on an FPGA as well as intra-pipeline
parallelism. A single memory feeds the pipelines and acts as a sink
for the results.
This FPGA design provides better performance than the CPU if the
added parallelism of these replicated pipelines can exceed the CPU's
increased clock rate and parallelism from instruction-level
parallelism and multiple cores. A very high effective memory bandwidth
for matrix elements is required because the pipelines' data originate
and terminate from the same memory component as shown in
Figure~\ref{fig:spgemmarchdesign}.

\textbf{Regularizing the FPGA data access.}  Keeping the pipelines
busy is the key to performance. A traditional problem with sparse
computations is the number of indirections needed to access the matrix
elements. When accesses are regular the FPGA can provide significant
acceleration for computation compared to a CPU because it can execute
thousands of floating point operations using digital signal processing
(DSP) units with support for fused-multiply-add operations, and its
high throughput distributed on-chip memory can store intermediate
results, thus avoiding write-backs to DRAM.

However, common sparse formats (\eg, CSR) involve some level of
indirection. For example, to access an element $A[i,j]$ in the CSR
format of the sparse matrix $A$, one needs to consult the row pointer
to obtain the location where the \textit{row i} begins (\ie,
$row\_pointer[i]$), and then search the column indices until the
beginning of the next row to check if the item is present(\ie, search
from $col[row\_pointer[i]]$ to $col[row\_pointer[i+1]$), and then
  subsequently access the data element at the matched index.

This irregularity in the computation that involves indirect accesses
prevents effective pipelining of operations resulting in low
throughput and resource utilization on the FPGA. Our design uses the
CPU to both schedule and pack the matrix elements in an intermediate
representation that increases regularity and resource utilization when
computed on the FPGA.

\begin{figure}[t]
\begin{center}
  \includegraphics[width=0.45\textwidth,angle=0]{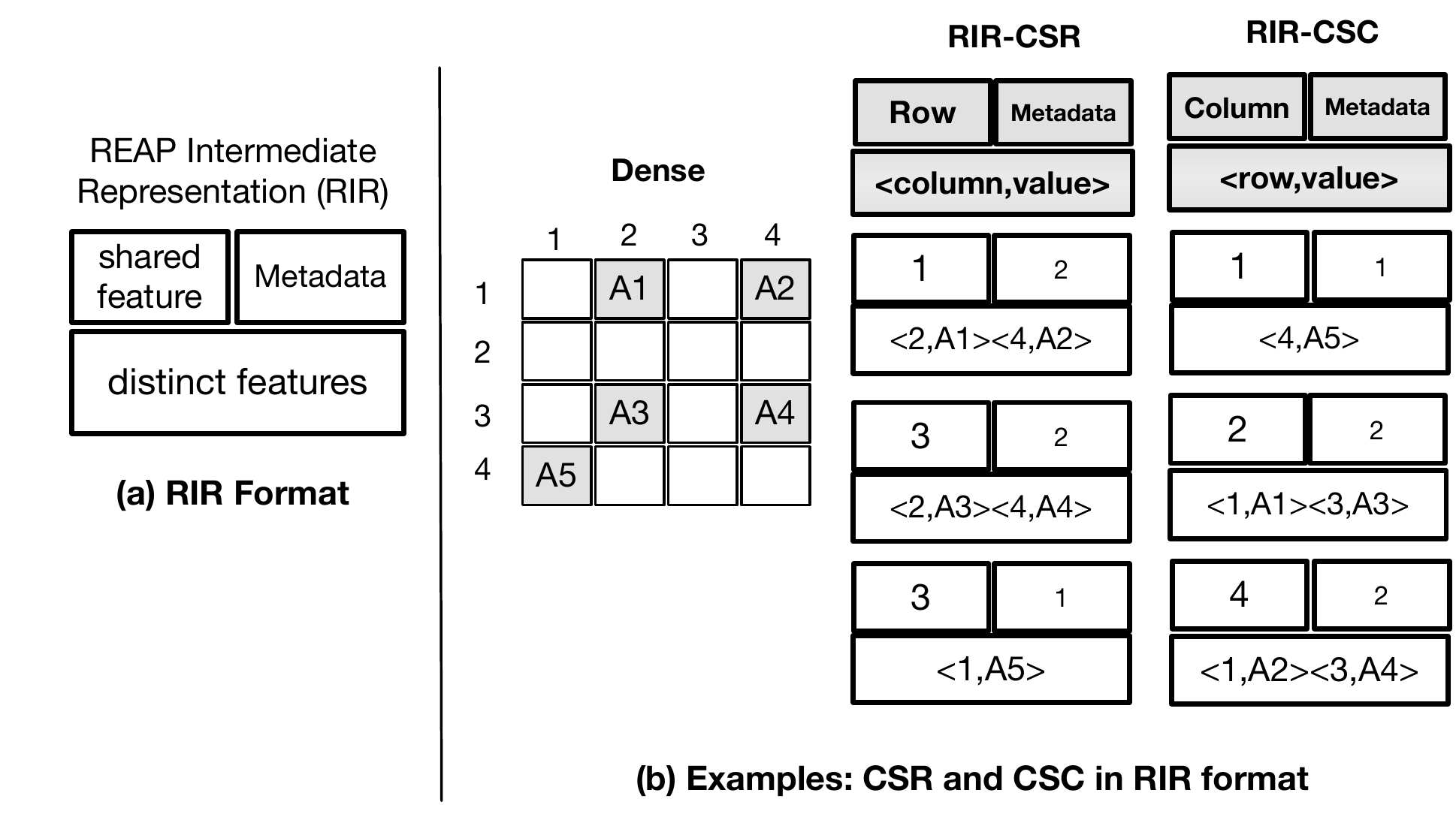}
  \caption{REAP's intermediate representation for increasing the
    throughput on FPGAs. (a) General RIR format. (b) Examples of RIR
    bundles for CSR and CSC sparse formats.}
\label{fig:ir}
\end{center}
\end{figure}

\textbf{REAP's intermediate sparse representation.}  REAP Intermediate
Representation (RIR) increases locality and minimizes indirect
accesses.  RIR co-locates both the value and the auxiliary
indices/metadata based on the shared feature. The key idea behind many
sparse formats including Compressed Sparse Format (CSR) is to group
non-zero elements based on some shared feature, which allows compact
storage of distinct features.  RIR is inspired by the very same
observation. It is straightforward to convert other sparse formats
such as CSC, ELL, and diagonal formats to RIR.

To provide FPGA's pipelines with a regular data stream, REAP performs
a data pre-processing pass using the software on the CPU.  In
Figure~\ref{fig:spgemmarchdesign}, the CPU takes the matrices from its
DRAM memory (not shown) and reformats them into RIR format placing
them in the FPGA memory. The CPU also uses the information from this
pre-processing pass to schedule the computation on the FPGA. The FPGA
then stores the results back in its own DRAM, which the CPU can
subsequently read them.

Figure~\ref{fig:ir}(a) depicts the RIR format where the shared feature
is the common attribute to all the distinct elements. For example, in
the case of a CSR representation, all the elements belonging to a
particular row will be packed together where the shared feature is the
row index. The column index and the value are also co-located with each
shared feature.

Figure~\ref{fig:ir}(b) provides the translation of commonly used
formats, CSR and CSC, to RIR, respectively.

One key feature of RIR
is that it allows easy conversion to commonly used formats. To support
any sparse format, one has to provide compress and decompress
routines. The compress routine takes the sparse data in the format of
interest and converts it to RIR.  Similarly, the decompress routine
converts from the data in RIR to a format of interest. This
reformatting is all done in software by the CPU. This is another
important advantage of RIR is to keep the FPGA design independent of
the sparse format while providing good performance.

\textbf{RIR includes scheduling.} RIR bundles can sometimes carry
purely the scheduling information rather than data. This scheduling
information guides the FPGA to schedule the data for the computation
and organize the data in the memory. This is especially useful in
sparse linear algebra kernels such as sparse Cholesky factorization
with data dependency. In other words, in those kernels, the output of
the computation can be served as the input in the later rounds. It is
inefficient to send the outputs to the CPU and read them back
later. Instead, in those cases, the CPU performs symbolic analysis and
uses it to guide the FPGA to organize the data in the memory without
the need to transfer any data between the CPU and FPGA.

Fine-grained communication with the CPU and the FPGA can be expensive
especially when the communication happens over IO channels (\eg,
PCI-E)~\cite{choi:quantitaveCPUFPGA:dac:2016}. The use of RIR with
coarse-grained scheduling information minimizes this overhead. In
contrast to executing everything on the FPGA, this synergistic
CPU-FPGA approach that makes two passes over the data enhances
locality, increases memory bandwidth, and provides higher throughput.

In the next section, we show how we adopt this approach and specific
design details for two important sparse kernels, sparse matrix
multiplication, and sparse Cholesky factorization. Similarly, many
other sparse linear algebra kernels can be accelerated with the same
approach, which we leave them for future work.

\section{Case studies}

To better show how REAP can be applied in practice, we present a
detailed design for two important sparse linear algebra kernels,
sparse generalized matrix multiplication and, sparse Cholesky
factorization. We chose these two kernels to study because they are
core components of many applications, such as graph
traversal~\cite{blas}, Monte-Carlo simulations~\cite{montecarlo}, and
Kalman filters~\cite{kalmanfilter}. Further, SpGEMM and Cholesky are
also different in their degrees of parallelism. Our goal is to
demonstrate that our cooperative technique applies to both strongly
parallel kernels and ones with data-dependencies.

\subsection{Accelerating SPGEMM with REAP}
\label{spgemm-detail}

A SpGEMM kernel consists of two main tasks: multiply and merge (also
known as accumulation). The multiply task generates partial
products. A partial product is a result of multiplying a non-zero
value of input A with a matched non-zero value of input B. A match
occurs when the column index of A matches the row index of B. The
merge task accumulates all partial products that have the same
coordinates (\ie, row and column) and produces the final result. Some
SpGEMM algorithms offer good throughput for the multiply task but have
difficulty with the merge task~\cite{pal:outerspace}. It is necessary to
balance the work between the multiply and merge tasks to achieve high
throughput.

We use a row-by-row formulation of SpGEMM that computes one row of the
result matrix at any instant of time. This formulation has two
advantages: (1) it provides improved data reuse compared to
inner-product and (2) it ensures lower complexity in the maintenance
of partial products compared to the outer-product
approach~\cite{pal:outerspace}.

\begin{small}
\begin{algorithm}
    \caption{A row-by-row formulation of SpGEMM}
    \label{alg:spgemmrowbyrow}
    $procedure \hspace{0.1cm} \textbf{SpGEMM}(Input A,B, Output C) $ \\
    \For{\texttt{each rowA $\in$ A}}{
	$NZA  = GetNonZero(A,rowA)$ \\
	$ROWB \gets \emptyset $ \\
        \For{\texttt{each (colA,valA) $\in$ NZA}}{
            $ROWB \gets  ROWB \cup \{colA\}  $\\
        }
        \For{\texttt{each rowB $\in$ ROWB}}{
            $NZB \hspace{0.1cm} = GetNonZero(B,rowB) $ \\
	    $(c_A,v_A) \hspace{0.1cm} \gets \{ (c_i,v_i) \in NZA \mid c_i=rowB \} $ \\
            \For{\texttt{each (colB,valB) $\in$ NZB}}{
		$valP\hspace{0.1cm} = valB \times v_A $\\
		$COLB \gets  COLB \cup \{colB\}  $\\	
                $PP \gets PP \cup \{(colB,valP)\} $\\
            }
        }
        $ sort (PP) $\\
	\For{\texttt{each colB $\in$ COLB}}{ 
		$M \hspace{0.1cm} \gets \{ (c_i,v_i) \in PP \mid c_i=colB \} $ \\	
		$y \hspace{0.1cm} = \sum_{(c_i,v_i)\in M}^{} v_i $ \\
		$C \gets C \cup \{(rowA,colB,y)\} $ \\
	}  
    }
\end{algorithm}
\end{small}

\textbf{Reducing storage for partial products.}
Algorithm~\ref{alg:spgemmrowbyrow} illustrates our row-by-row SpGEMM
formulation for multiplying two sparse matrices $A$ and $B$.
Intuitively, each row of $A$ is compared with all the rows of $B$,
matching elements (\ie, when the column index of an element in $A$ and
the row index of an element in $B$ are equal) are multiplied to
generate partial products, and the partial product belonging to the
same column index are accumulated to produce a row of the final result
matrix.  The $A$-matrix is read once and the $B$-matrix is streamed
into the FPGA for each row of $A$.

Given that matrices $A$ and $B$ are sparse, it is not necessary to
stream all the rows of $B$ for a given row of $A$. When we read a row
of $A$, we identify the column indices of the non-zero elements in
that row and only stream those rows of $B$ that matches one of the
column indices of $A$ (lines 3-7 in
Algorithm~\ref{alg:spgemmrowbyrow}). For example, if there is only one
non-zero element in a row of $A$, then it is not necessary to stream
all rows of $B$. Just streaming one row of $B$ that matches the column
index of the single non-zero element of $A$ is sufficient.

Given a row of $A$ and a row of $B$, the algorithm identifies a
non-zero element of $A$ in that row which matches with the elements in
a row of $B$ and performs the multiplication~(lines 10-14 in
Algorithm~\ref{alg:spgemmrowbyrow}).  The multiplication of matching
elements produces a partial product. Each partial product maintains
the value and column index of the result. After streaming all the rows
of $B$, the partial products are sorted and merged by accumulating the
values with the same column index. The merged values after
accumulation provide the non-zero elements of a row in the final
result matrix~(lines 17-21 in Algorithm~\ref{alg:spgemmrowbyrow}).

\begin{figure}[t]
\begin{center}

\includegraphics[width=0.5\textwidth,angle=0]{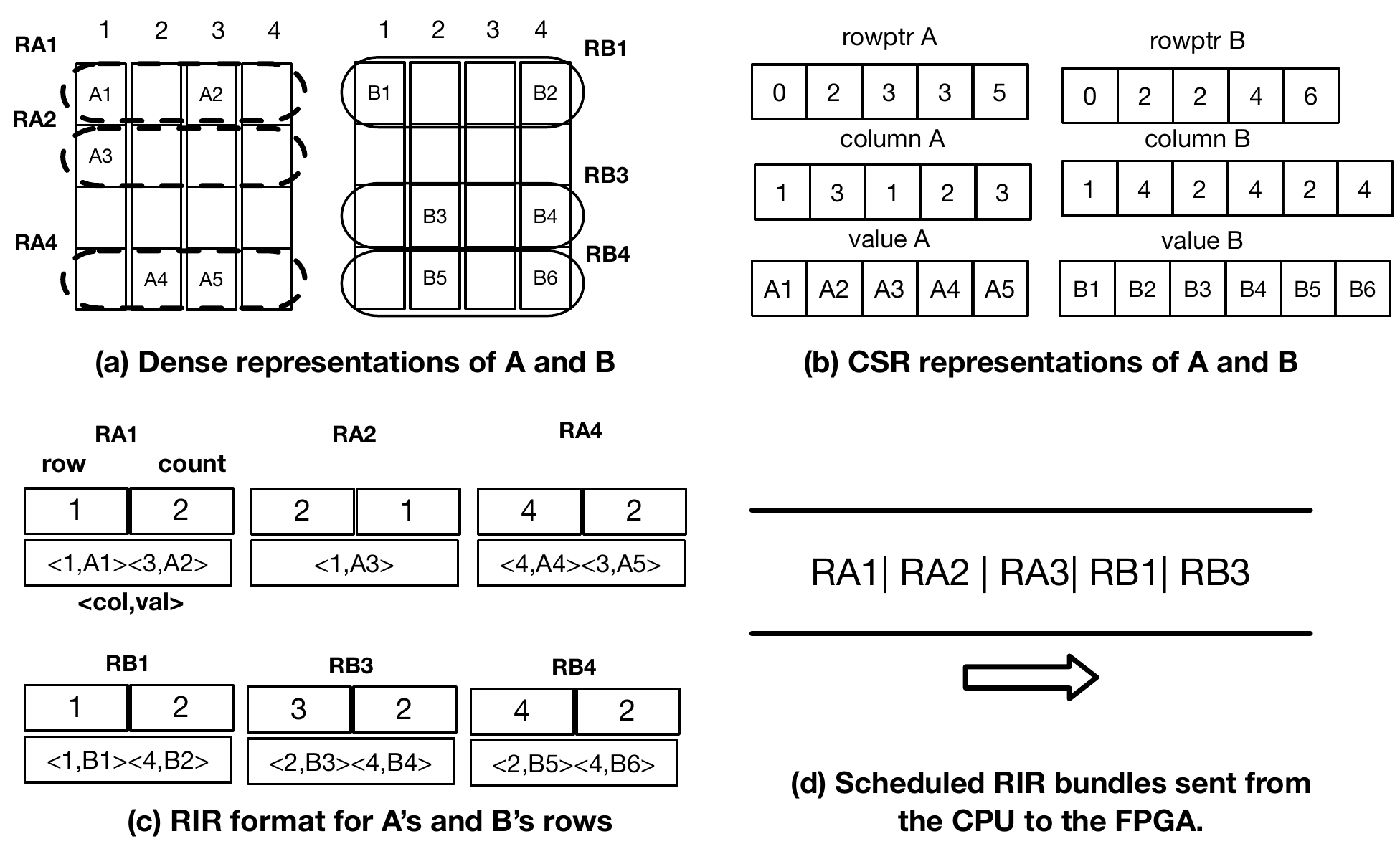}
\caption{\small Illustration of CPU's activities to reorganize the
  data and to provide scheduling information. (a) Example sparse
  matrices $A$ and $B$ shown in dense representation. (b) CSR
  representation of the two matrices. (c) The data reorganized by the
  CPU as RIR bundles where the value and column indices are co-located
  with each shared feature. (d) The layout of the RIR bundles in
  memory provided by the CPU to the FPGA.}
  \label{fig:spgemmarch}
\end{center}
\end{figure}

Hence, the main components to accelerate SpGEMM in our formulation
involve (1) extraction of the non-zero elements of two matrices in a
given row and (2) efficient execution of the pipeline consisting of
multiply, sort, and merge operations. In our synergistic CPU-FPGA
approach, the CPU performs the first task above whereas the FPGA
performs the second task.

\textbf{CPU provides regular data and scheduling information in
  the RIR format.}  CPU reorganizes the input data to make it easier
for the FPGA to attain higher throughput. CPU has information about
the FPGA design (\ie, number of pipelines and the PE's in each unit)
and uses it to layout the data. CPU reads both sparse matrices in CSR format and
creates RIR bundles for each row of the input matrices. During this
process, it also layouts the bundles in memory (DRAM) by using the
scheduling information.

Figure~\ref{fig:spgemmarch} illustrates how the data is organized by
the CPU to make it regular and to encode the scheduling information
for the FPGA. CPU is aware of the number of parallel pipelines in the
FPGA (\ie, three in Figure~\ref{fig:spgemmarchdesign}) to properly
perform the scheduling task. Each pipeline processes a row of
A. Hence, it has laid out the three rows of A followed by all the rows
of B necessary to produce all partial products.  In summary, CPU and
FPGA work in tandem to attain higher throughput.

\textbf{FPGA design for SpGEMM.} Figure \ref{fig:spgemmarchdesign}
shows the details of the architecture of our SpGEMM design.  The FPGA
design consists of five modules that process the input data
in the RIR format.  The input controller reads the input matrices and
the scheduling information in RIR and distributes the work
to the pipelines. Each pipeline includes PEs that perform match and
multiply unit, sorting, and merging. Distinct pipelines operate on a
distinct row of matrix $A$ and all rows of $B$ are streamed to every
pipeline. Each pipeline computes a small part of the matrix independently.

\textbf{Match and multiply unit.} The match and multiply unit consists of two smaller units - a match
unit and a single-precision multiplier. We use Content Addressable
Memory (CAM) to perform matching. Each CAM is populated with columns
of A as the key and the address to another location in memory where
A's row and value are stored as the value. The match and multiply
units are connected by a buffer that acts as the work queue for the
multiplier unit. Once, a column index of an element in the row of A
matches with the row index of an element of B's row, the values of the
two elements of A and B are pushed to the multiplier work-queue. The
multipliers produce partial products. 

\textbf{Sorting of partial products.} These partial products then
sorted by the sort unit before being merged. To design a high-speed
sorting unit, we use shift registers and attach comparators to each
register. Initially, all the entries in the shift register are
invalid. When a partial product is created, we want to find where the
new element should be inserted. The sorting unit compares the new
value (\ie, the column index of the partial product) with all the
existing values, identifies the appropriate position to insert, and
shift other values to make space if necessary.

\textbf{Merging of partial products.} The merge units accumulate all
the incoming partial products that have the same column index.  The
partial products are kept in a queue for the merge. As all the partial
products belong to the same row in the pipeline and they are sorted
according to their column index, the new element for the merge is just
compared with the top element. If the new element has the same column
index as the top element in the queue, it is accumulated. Otherwise,
the top element is removed and sent to the output controller and the
new element is added to the queue.

\textbf{Improving scalability.} Our goal is to handle matrices of
any size with any sparsity pattern. Some rows can have a large number
of non-zeros compared to the number of resources in a pipeline in our
FPGA design. In such scenarios, it is necessary to split the row into
smaller pieces. RIR format encodes the number of elements in each bundle. 
This allows us to break a large row into smaller RIR bundles.  When the CPU
packs the input data into an RIR bundle, it encodes the limit on the
number of elements in it, which is a design parameter.  Our match and
multiply units in the FPGA design use CAMs.  An increase in the number
of CAMs impacts the frequency of the design. In our SpGEMM design, we
use an RIR bundle size of 32. When the number of non-zero elements in a
row exceeds the RIR bundle size, CPU breaks the whole row into
multiple bundles. The RIR bundle also includes additional metadata to
indicate the end of a row in the input matrix. This organization of
the data and the scheduling information provided by the CPU enables
the input and join controllers of the FPGA design to work efficiently.

\textbf{Summary.} To maximize the throughput, it is essential to keep
all units in a pipeline and all pipelines active. The organization of
the data and the schedule by the CPU minimizes the complexity of the
input controller that its job is to distribute the inputs and enables
effective utilization of the pipelines and the PEs in the FPGA.

\subsection{Accelerating Cholesky with REAP}
\label{sec:cholesky}
Cholesky factorization~\cite{Heath:ScientificComputing} is an
important method to solve systems of equations, $Ax=b$. If the matrix
is positive semi-definite, and thus symmetric, then we can decompose
the matrix $A$ into $LL^{T}$. The Cholesky method computes the lower
triangular $L$ matrix from $A$.  In the left-looking Cholesky, the
columns of L are computed from left to right. The two main challenges
of a sparse Cholesky factorization are:
\begin{itemize}
\item There are data dependencies between the computation of columns
  of L.  We cannot begin the computation of column $K$ until all the
  dependencies are resolved.  This is in contrast to SpGEMM, where
  each row of the result matrix could be computed independently and in
  any order.
\item The non-zero structure of the sparse matrix L is modified as we
  compute L. In other words, L is both an input and an output.  This
  makes Cholesky more challenging compared to SpGEMM where the
  sparsity pattern is fixed before the computation starts.
\end{itemize}

\begin{small}
\begin{algorithm}
    $procedure \hspace{0.1cm} \textbf{SparseCholesky}(Input A, Output L) $ \\

    \For{\texttt{each column k in $\in$ A}}{ 
	
       $DOT = GetNonZeroCol(A,k)$ \\
       $ROWL = GetPattern(L,k) $ \\
    	\For{\texttt{each row r $\in$ ROWL}}{
        	$DOT(r) \hspace{0.15cm} -= L(r,0:k-1) \hspace{0.15cm} . \hspace{0.15cm} L(k,0:k-1) $ \\
    	}
 
    	$L(k,k) = sqrt(DOT(k)) \hspace{0.1cm} \triangleright $ \texttt{Diagonal} \\
    	\For{(row,value) $\in$ DOT}{ 
        	$L(r,k) = value \hspace{0.15cm} / \hspace{0.15cm} L(k,k) \hspace{0.15cm} \triangleright $ \texttt{Off-Diagonal}\\ 
    	}
    }
    \caption{The left-looking Cholesky factorization.}
    \label{fig:leftcholalg1}
\end{algorithm}
\end{small}

\textbf{Computing Cholesky factorization.}
Algorithm~\ref{fig:leftcholalg1} shows the steps in a left-looking
Cholesky factorization. The matrix $L$ is computed column-by-column
based on the column of A and the previous column of L in the outer-loop
beginning in line 2. To compute the $k$-th column of L, one needs the
$k$-th column of A and all the columns of $L$ from $[0, k-1]$ and all
rows of $L$ starting from row $k$. Given that $A$ is sparse, we
first read the non-zero elements in a given column of $A$ (line 3).
The column vector {\tt DOT} holds partial results. It is initialized
to be equal to the non-zero elements of the column in $A$.

An interesting aspect of Cholesky factorization is that it is possible
to identify the non-zero elements in a column of $L$ from a pure
symbolic analysis of the dependencies of $L$ and the values of
$A$~\cite{directmethod:timothy}. Hence, the algorithm performs a
symbolic analysis to identify the row indices that are non-zeros in a
column of $L$ (\ie, \texttt{ROWL} in line 4 in
Algorithm~\ref{fig:leftcholalg1}).
To compute a non-zero element in column $k$ with a row index $r$, we
need to compute the dot product of the $r$-th row and $k$-th row of
$L$ which is then subtracted from the values in $A$'s $k$-th column
with row index $r$. \texttt{DOT} stores this partial value (lines
5-7).  Finally, the diagonal and non-diagonal elements of the $k$-th
column of $L$ are computed as shown~(lines 8-11 in
Algorithm~\ref{fig:leftcholalg1}).

\begin{figure}[th]
  \centering
  \includegraphics[width=0.49\textwidth]{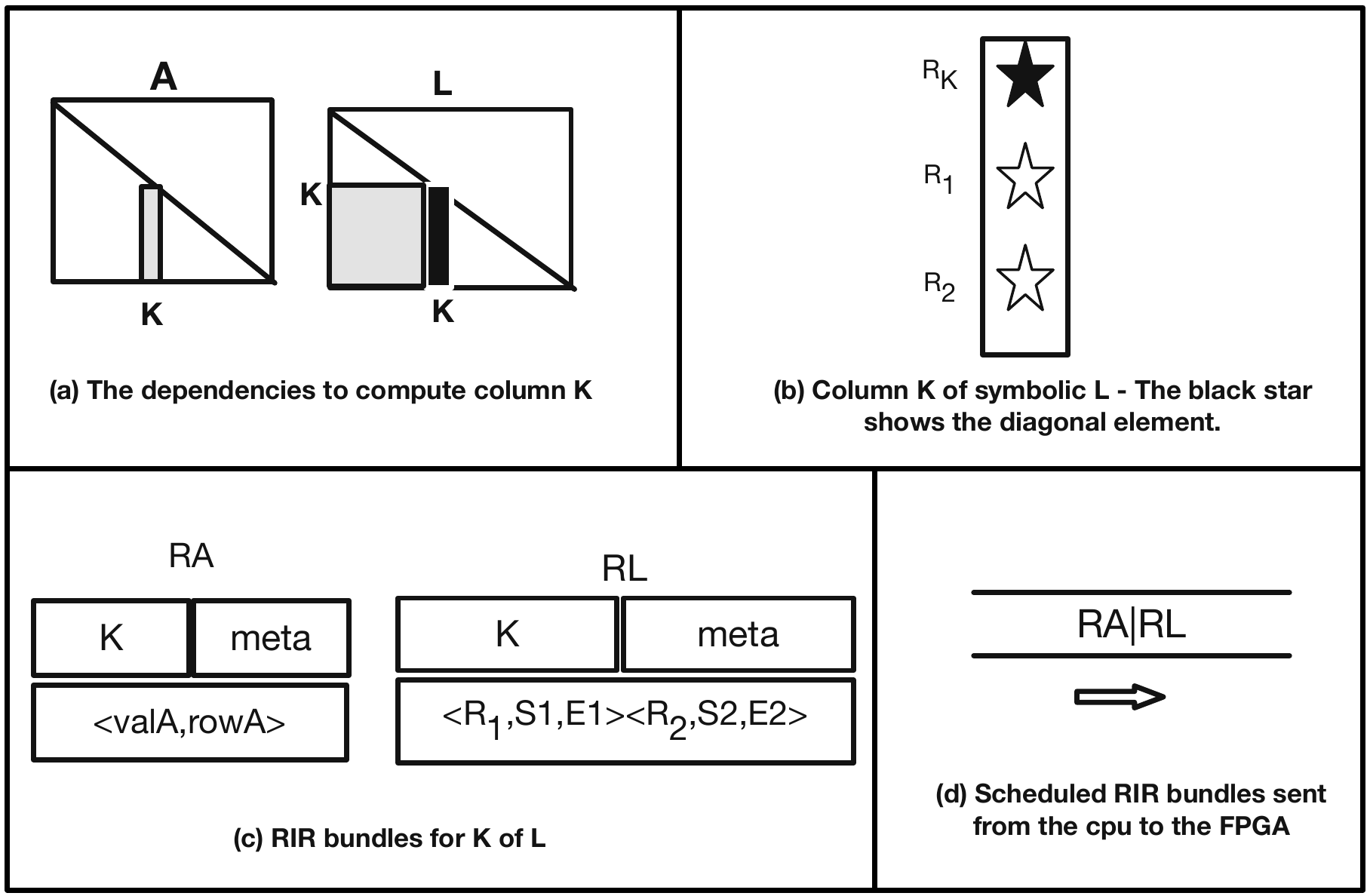}
  \caption{\small CPU's work in creating RIR bundles and symbolic
    analyses with Cholesky.  (a) The dependencies for computing a
    column $K$ of matrix $L$ are shown in gray. It needs the $K$-th
    column of A and all the columns of $L$ from $[0, K-1]$ and all
    rows starting from $K$. (b) The symbolic analysis on the CPU
    generates the set of elements in the column $K$ of matrix $L$ that
    are going to be non-zero without actually doing any numeric
    computation. (c) RIR bundles generated by the CPU to provide
    information about scheduling. As $L$ resides in FPGA's memory, the
    CPU also provides information about where a particular row R1 of
    $L$ starts and ends (\ie, S1 and E1). (d) The regular data sent by
    the CPU to the FPGA.}
  \label{fig:chol_1}  
\end{figure}

\textbf{Challenges for FPGA acceleration.} Sparse Cholesky
factorization is challenging for FPGAs because the computation depends
on the sparsity pattern. The location of non-zeros in a column of $L$
depends on the value of $A$. FPGAs are pretty deficient in performing
irregular accesses that happen with symbolic analysis. In the absence
of such symbolic analysis, it is unclear how to schedule computation
efficiently on the FPGA.

\textbf{Symbolic analysis and data reorganization by the CPU.}  REAP's
synergistic CPU-FPGA acceleration addresses these issues and provides
acceleration for Cholesky factorization using FPGAs.  In our approach,
CPU performs the symbolic analysis based on the construction of the
\textit{elimination
  tree}~\cite{george1987symbolic,zhu1994constructing}.
CPU pre-processes the $A$ matrix and performs a symbolic analysis of
$L$ and $A$ to identify non-zeros rows in any column of
$L$~(\texttt{GetPattern} in
Algorithm~\ref{fig:leftcholalg1}). Figure~\ref{fig:chol_1} (b) shows
the result of symbolic analysis indicating the rows that are non-zeros
in a column of L.

To regularize the data, CPU generates RIR bundles for $A$.
Figure~\ref{fig:chol_1}(c) shows the RIR bundles for matrix $A$, which
is labeled RA, where k is the column index (shared feature) and row
and value pairs are the distinct features of the bundle.
CPU also creates metadata RIR bundles to convey the scheduling
information for computing a column of $L$. As $L$ matrix is
exclusively allocated and stored in the FPGA, CPU determines the size
of the $L$ matrix and creates a metadata RIR bundle for $L$.  The
metadata bundle for column $K$ is a vector of triples, which is
labeled RL in Figure~\ref{fig:chol_1}(c). It includes the row index
($r$) of the non-zero value in a column $K$ of $L$, the start and end
indices of the $r$-th row of L.
Overall, CPU performs lines 3-4 in Algorithm~\ref{fig:leftcholalg1}
and the FPGA performs the rest.

\begin{figure}[th]
  \centering
  \includegraphics[width=0.49\textwidth]{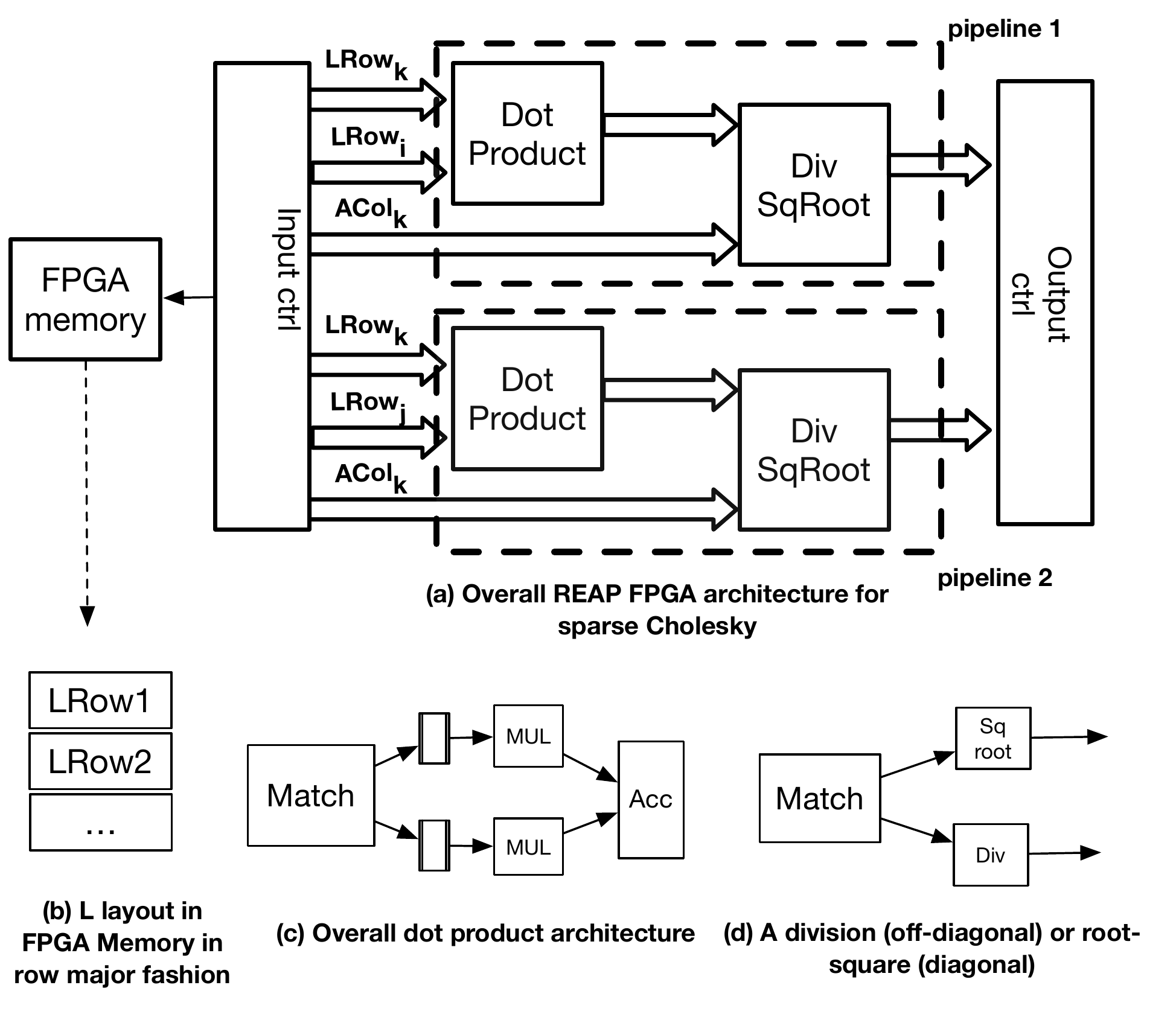}
  \caption{Sparse Cholesky FPGA Architecture}
  \label{fig:chol_arch}  
\end{figure}

\textbf{Acceleration of computation in the FPGA.}  Figure
\ref{fig:chol_arch} shows the overall architecture for the computation
of Cholesky factorization on the FPGA. It computes a column of $L$ in
parallel. The FPGA design consists of a collection of pipelines. Each
pipeline computes a non-zero element of a column of $L$. Each pipeline
internally contains a pipeline of processing elements that perform the
dot product and a division or a square root operation.

FPGA's input controller uses the RIR bundles for $A$ and the metadata
bundles for $L$ to distribute the computation. The matrix $L$ is
maintained in the memory accessible by the FPGA. While computing
column $k$ of matrix $L$, the input controller broadcasts (1) row $k$
of matrix $L$ and (2) RIR bundle for column $k$ of $A$ to all the
pipelines. Additionally, each pipeline retrieves a unique row of $L$
corresponding to a non-zero element in column $k$ of $L$.

Each dot product unit in the pipeline is equipped with a CAM to match
the indices and several multipliers as shown in
Figure~\ref{fig:chol_arch}(c). The design is fully pipelined by adding
intermediate buffers between each component of the design. Each dot
product PE has multiple multipliers to increase parallelism within the
PE.
The results of the dot products are then used by the second kind of
PEs, labeled ``Div SqRoot'', to compute the final values for each row
of a column of $L$, as shown in Figure~\ref{fig:chol_arch}(d).  This
PE realizes lines 8-11 in Algorithm~\ref{fig:leftcholalg1}.

The non-diagonal elements of a column of $L$ also depend on the value
of the diagonal element. To make the computation of each pipeline
completely independent, each pipeline computes the diagonal element
independently and increases throughput at the cost of performing some
redundant computation.

\textbf{Summary.} CPU performs the symbolic analysis, regularizes the
data, and identifies the scheduling of computation that enables the
FPGA design to concurrently compute the entire column of $L$ in
parallel and attain higher throughput.

\section{FPGA Prototype}

We realized REAP prototypes of SPGEMM in 12,424 lines of hand-coded
Verilog and Cholesky in 9072 lines, including the test-benches. We
synthesized these codes for an Intel Altera DE5net-Arria board, using
Quartus-16.1. The rest of this section describes some of the design
choices.

\textbf{Reading and writing RIR bundles.} We used Intel's FIFO block
in the Quartus development environment and designed read/write
controllers to access RIR bundles. We used dual-ported FIFOs, but the
read and write signals synchronized to the same clock.

The write controller performs the following tasks: (1) It writes all
the distinct features and keeps track of the number of elements
written to the FIFO. (2) Once all the distinct features are written,
it pushes the shared feature and the metadata that provides the number
of distinct elements written. The read controller reads the bundle in
the reverse order. It reads the metadata first, shared feature next,
and finally the distinct elements of the bundle.  To handle overflow
with FIFOs, the controllers make use of an almost-full signal (with
writes) and an empty signal (with reads).

\textbf{Floating Point Operations.} All the arithmetic computation
including multiplication, addition, subtraction, division, and square
root operations, in both designs, use dedicated hardware (i.e. from
the DSP units) for single-precision floating point. We did not use
double-precision as there is no support for it as vendor-supplied
Intellectual Property (IP) blocks. We did not explore synthesizing
double-precision FP arithmetic using FPGA logic cells as this has been
shown to have much worse frequency and area than dedicated FP
units~\cite{beauchamp2008architectural}.

\textbf{REAP with High-Level Synthesis.} As high-level synthesis (HLS)
is becoming an attractive alternative to designs with hand-coded
Verilog, we explored ideas from REAP with OpenCL HLS tools for
FPGA. We observed that the performance of the HLS designs tend to be
lot slower compared to hand-coded Verilog and can vary with minor
changes to the design. In the evaluation, we demonstrate that ideas
from REAP such working on pre-processed data can speedup HLS designs.

\section{Experimental Evaluation}
\label{evaluation}

\begin{table}[t]
  \centering
  \scriptsize
  \begin{tabular}{|c|c|c|c|c|}
    \hline
    Name & SpGEMM & Cholesky & Row & NNZ(Density) \\\hline
    mario\_002 & S1 & - & 389K & 2.10M(0.001\%) \\
    m133-b3 & S2 & - & 200K & 800K(0.001\%) \\
    filter3D & S3 & - & 106K & 2.7M(0.02\%) \\
    cop20K & S4 & - & 121K & 2.6M(0.01\%) \\
    offshore & S5 & - & 259K & 4.2M(0.006) \\
    poission3Da & S6 & - & 13K & 352K(0.19\%) \\
    cage12 & S7 & - & 130K & 2.0M(0.011\%) \\
    2cubes\_sphere & S8 & - & 101K & 1.64M (0.015\%) \\
    bcsstk13 & S9 & C2 & 2K & 83K(2.09\%) \\
    bcsstk17 & S10 & C3 & 10K & 428K(0.35\%) \\
    cant & S11 & C4 & 62K & 4M(0.102\%)\\
    consph & S12 & - & 83K & 6M(0.086\%)\\
    mbeacxc & S13 & - & 496 & 49K(20.29\%) \\
    pdb1HYs & S14 & - & 36K & 4.3M(0.32\%)\\
    rma10 & S15 & - & 46K & 2.3M(0.108\%)\\
    descriptor\_xingo6u & S16 & - & 20k & 73k(0.017\%) \\
    g7jac060sc & S17 & - & 17k & 203k(0.064\%) \\
    ns3Da & S18 & - & 20k & 1.6M (0.403\%)\\
    TSOPF\_RS\_b162\_c3 & S19 & - & 15k & 610k(0.25\%) \\
    cbuckle & S20 & C6 & 13k & 676k(0.36\%)\\
    Pre\_poisson & - & C1 & 12K & 715K(0.32\%)\\ 
    gyro & - & C5 & 17K & 1M(0.33\%)\\
    bcsstk18 & - & C7 & 11K & 80K(0.056\%) \\
    bcsstk36 & - & C8 & 23K & 1.1M(0.215\%)\\
    \hline
  \end{tabular}
  \caption{\small Matrices from SparseSuite
    \cite{davis:floridabenchmark} used in our evaluation. We use the
    second and third column as the ID to refer to the benchmarks when
    presenting the results.}
  \label{tbl:benchmarks} 
  \vspace{-4mm}
\end{table}

\begin{table}[t]
  \centering
  \scriptsize
  \begin{tabular}{|c|c|}
    \hline
    Platform & Configuration \\ \hline
    CPU                & 16 cores, 2.1 GHz, 32 GB DDR4 (2666 Mhz) \\ 
    Intel Xeon 6130    & Cache(KB) L1:32 L2:1024 L3:22528   \\ \hline
    FPGA & 1,150K logic elements \\ 
    DE5net- Arria 10& 67-Mbits embedded memory, 8 GB DDR3 (933 Mhz) \\ 
    & 1518 DSP blocks \\   \hline 
    Intel PAC Card  & 1,150K logic elements, \\ Arria 10 GX FPGA & 65.7 Mbit on-chip memory,  
    OpenCL 1.0 \\ \hline
  \end{tabular}
  \caption{The CPU and FPGA configurations}
  \label{table:config}
  \vspace{-4mm}
\end{table}

We evaluated REAP by comparing the performance of two sparse kernels,
SpGEMM and Cholesky against two well-optimized CPU libraries, the
Intel MKL for SpGEMM~\cite{intelmkl} and CHOLMOD~\cite{CHOLMOD:2008}
for Cholesky.

\textbf{Benchmarks}. We used twenty-four sparse matrices from the
SuiteSparse matrix collection~\cite{davis:floridabenchmark}, which are
shown in Table~\ref{tbl:benchmarks}. These matrices span multiple
application domains and vary in size and sparsity patterns.
The second and third columns are the matrix ID names in the evaluation
results. The blank means the matrix was not evaluated for that
algorithm (\ie, it does not satisfy the precondition for Cholesky
factorization).

\begin{figure*}[th]
  \centering
  \includegraphics[width=1\textwidth,angle=0]{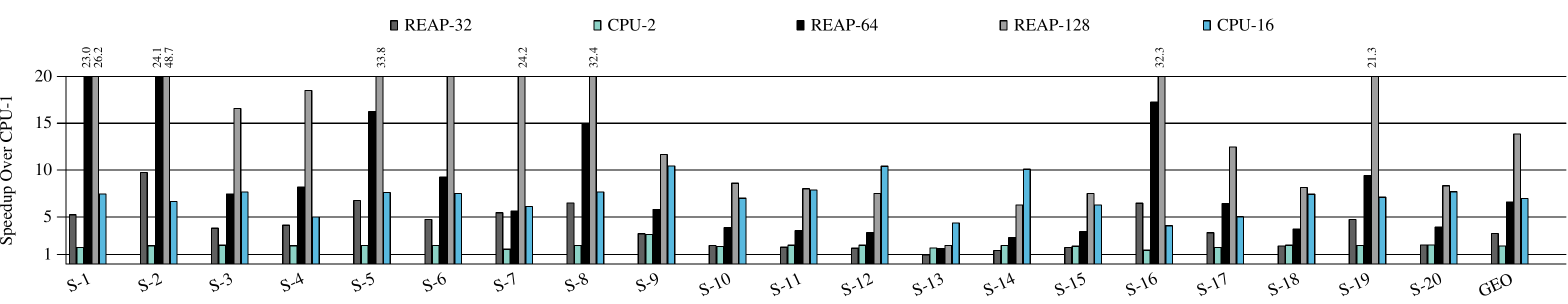}
\caption{\small Speedup of REAP designs and software-only
  multi-core CPU versions compared to Intel MKL on a single core for
  SpGEMM.}
\label{fig:spgemmspeedup}
\end{figure*}

\textbf{Simulation framework}. Although we have synthesized designs on
the FPGA, we are limited by the memory bandwidth for our experiments on the DE5-Net FPGA board. Recently, Intel ~\cite{intelhbm} and Xilinx ~\cite{xilinxhbm} released their newest high-end FPGAs with High Bandwidth Memory (HBM) interfaces using newer packaging techniques. As these boards
became recently available, we did not have access to them and
experiment on them. To explore the design space, we also use
simulation to measure the execution time of REAP designs when we scale the memory bandwidth. Our framework consists of two parts. First, for the FPGA
part, we use a trace-driven simulation with our in-house cycle-accurate SystemC
simulator for high fidelity. All the cycle counts and FPGA frequencies
used in the simulator are extracted from the RTL implementation
synthesized by Quartus 16.1 version for the DE5-Arria 10 board. The
details of the FPGA board are available in Table
\ref{table:config}. To simulate the FPGA DRAM, we use a queuing model
where the data transfers are not allowed to exceed the bandwidth set
in the design. Second, for the CPU part of REAP, we use a
single-thread C++ implementation. Our current framework does not
support concurrent execution of the CPU and FPGA. Hence, we initially
ran the CPU part to collect the CPU traces and then the simulator uses
the trace for its execution.

\textbf{Baseline}. We use the measured execution time of Intel's MKL's
SpGEMM implementation on a CPU as the baseline to compare with our
REAP SpGEMM designs. For sparse Cholesky, we measured CHOLMOD, a
specialized library for Cholesky factorization. Table
\ref{table:config} shows the specifics of the CPUs used and the
FPGA. To evaluate the SpGEMM kernel, we multiply each sparse matrix by
itself (\ie, $C=A^{2}$), which is a standard method for evaluating
SpGEMM performance.

\begin{figure}
\centering
 \includegraphics[width=0.50\textwidth,angle=0]{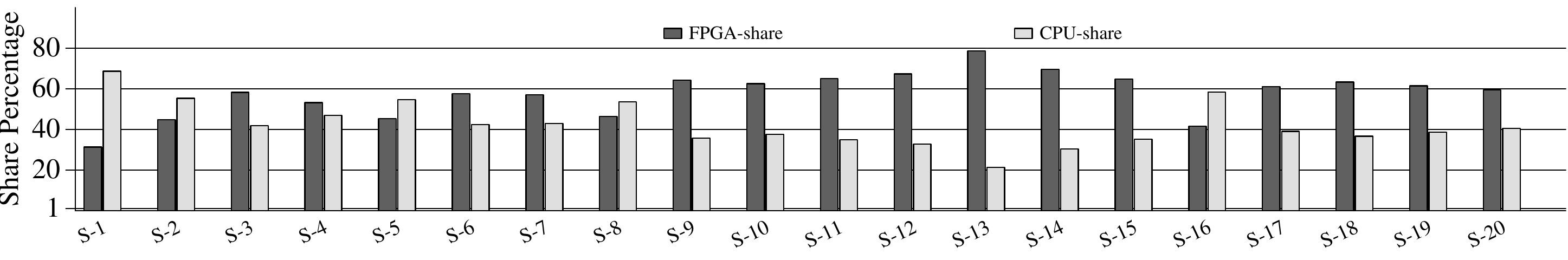}
 \caption{\small Percentage of the time spent in preprocessing with the CPU
   and in computation on the FPGA for our REAP-32 design for SpGEMM.}
 \label{fig:spgemm-prep}
\end{figure}


\subsection{Performance Evaluation with SpGEMM}
\label{spgemm_res}
We compare three variants of our SpGEMM design to Intel MKL.

\textbf{REAP-32}. This FPGA design has 32 pipelines and runs
at 250 MHz frequency. We use a RIR bundle and CAM size of 32.  The
DRAM bandwidth for this design matches that available on a single-core
CPU, which is 14 GB/s on our machine for both reads and writes. We
measured memory bandwidth with the {\tt pmbw} tool
\cite{pmbw:memorybandwith}.

\textbf{REAP-64}. This variant has 64 pipelines and runs at 250
MHz. The DRAM bandwidth is scaled to reflect the DRAM bandwidth
available for the 16-core CPU, which is the peak measured memory
bandwidth (147 GB/s for reads and 73 GB/s for writes) for our CPU. All
other parameters are same as the REAP-32 version.

\textbf{REAP-128}. This version has 128 pipelines and runs at 220
MHz. The DRAM bandwidth remains the same as \textit{REAP-64} which is
the peak memory bandwidth measured for our CPU system. All other
parameters are same as REAP-64.

\textbf{CPU.} We run Intel MKL with a different number of threads
(1,2, 4, 8, and 16), which indicates the change in performance on a
CPU with core count. We achieved the best performance for the CPU running the Intel MKL with 16 threads with hyperthreading disabled. Variants with more threads ran slower,
likely due to interference.

Figure \ref{fig:spgemmspeedup} shows the SpGEMM speedup of our REAP
FPGA designs and software-only CPU versions with various core counts
relative to Intel MKL run on a single core. REAP overlaps the reformatting on the CPU and the computation on
the FPGA after the initial round. In the initial round, the FPGA is
idle while CPU reformats the data.  Figure \ref{fig:spgemmspeedup} shows the overall time
taking into account both the CPU and the FPGA time.
Due to space reasons, we only show the least and the best performance
on the CPU and the three REAP variants. The CPU-2 effectively has the
same number of floating-point multiply/add units as the REAP-32 while
REAP-32 effective bandwidth is the same as the CPU-1. REAP-64 and
REAP-128 have respectively a quarter and half of the number of
floating-point multiply/add units than CPU-16, with the same memory
bandwidth.

The main observations from this figure are: (1) REAP-32, where the
FPGA has the same memory bandwidth as single-core Intel MKL
outperforms the CPU-1 on \textit{all} the matrices and with a
geometric mean of 3.2$\times$ speedup. (2) REAP-32 also beats CPU-2
for the majority of the benchmarks and REAP-64 outperforms CPU-16 for
half of the matrices, with 1/4 of the available floating points
units. This higher performance is achieved despite the FPGA running at
a frequency that is almost 8$\times$ slower than the CPU's
frequency. (3) REAP-128 beats CPU-16 for all benchmarks except
three. This is again achieved even when REAP is using half of the
number of floating-point units compared to a 16-core CPU and also
running with significantly lower frequency than the CPU.

\textbf{CPU preprocessing vs FPGA computation.}
Figure~\ref{fig:spgemm-prep} reports the percentage of the time spent
in CPU preprocessing task and FPGA computation with our REAP-32
design~(the sum of the two should add up to 100\%). In reality, most
of the execution times on CPU and FPGA are effectively overlapped.
From this figure, we observe that the FPGA computation time dominates
the CPU's preprocessing time for most cases, which is expected.  In
cases, where CPU preprocessing time exceeds FPGA's time, the matrices
have low density. In this scenario, the time spent to extract and
organize the non-zero elements is more than the computation time. This
confirms that accessing the non-zero elements could sometimes be
costly as doing the actual computation with SpGEMM.

\begin{figure}
\centering
\begin{minipage}{.25\textwidth}
  \centering
  \includegraphics[width=1\textwidth,angle=0]{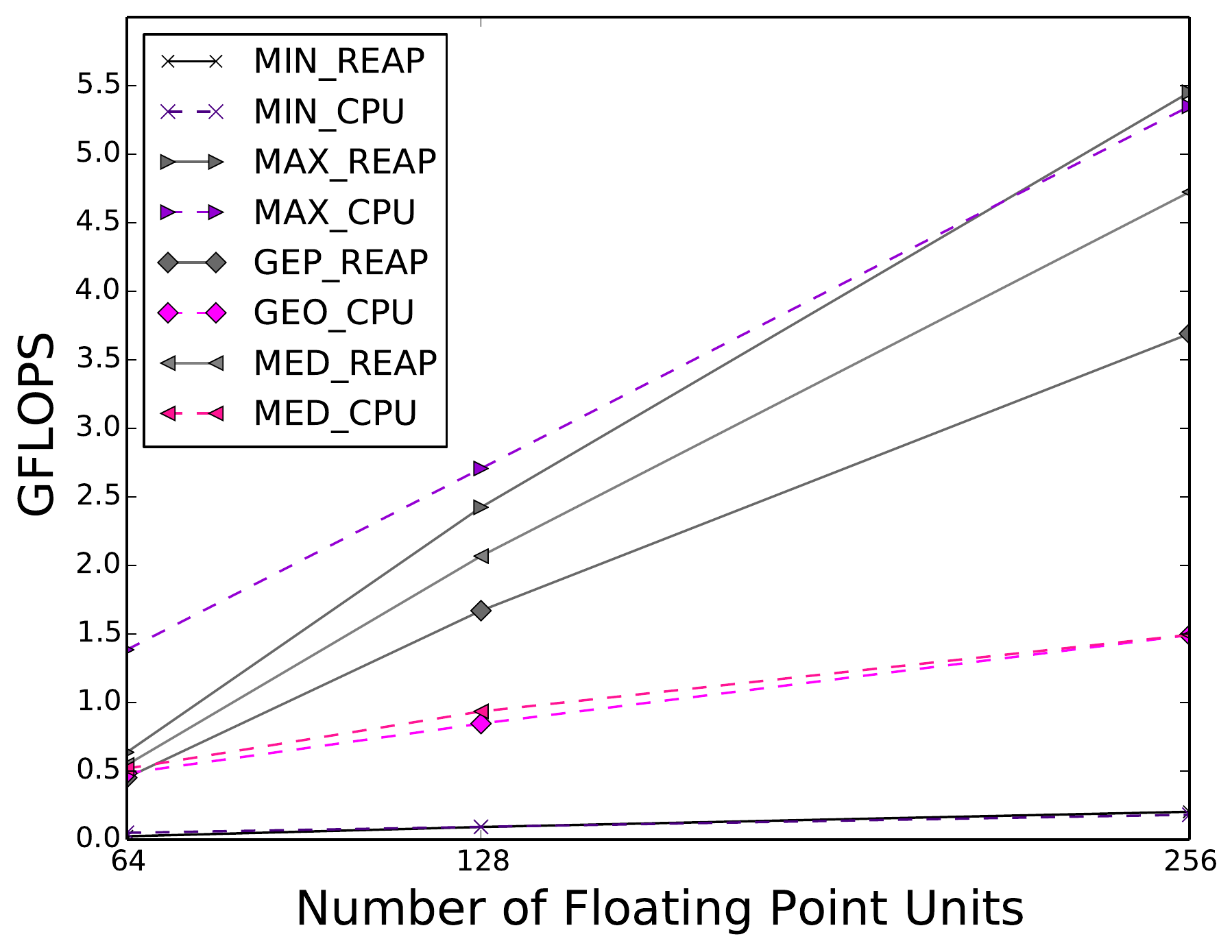}
  
\end{minipage}%
\begin{minipage}{.25\textwidth}
  \centering
  \includegraphics[width=1\textwidth,angle=0]{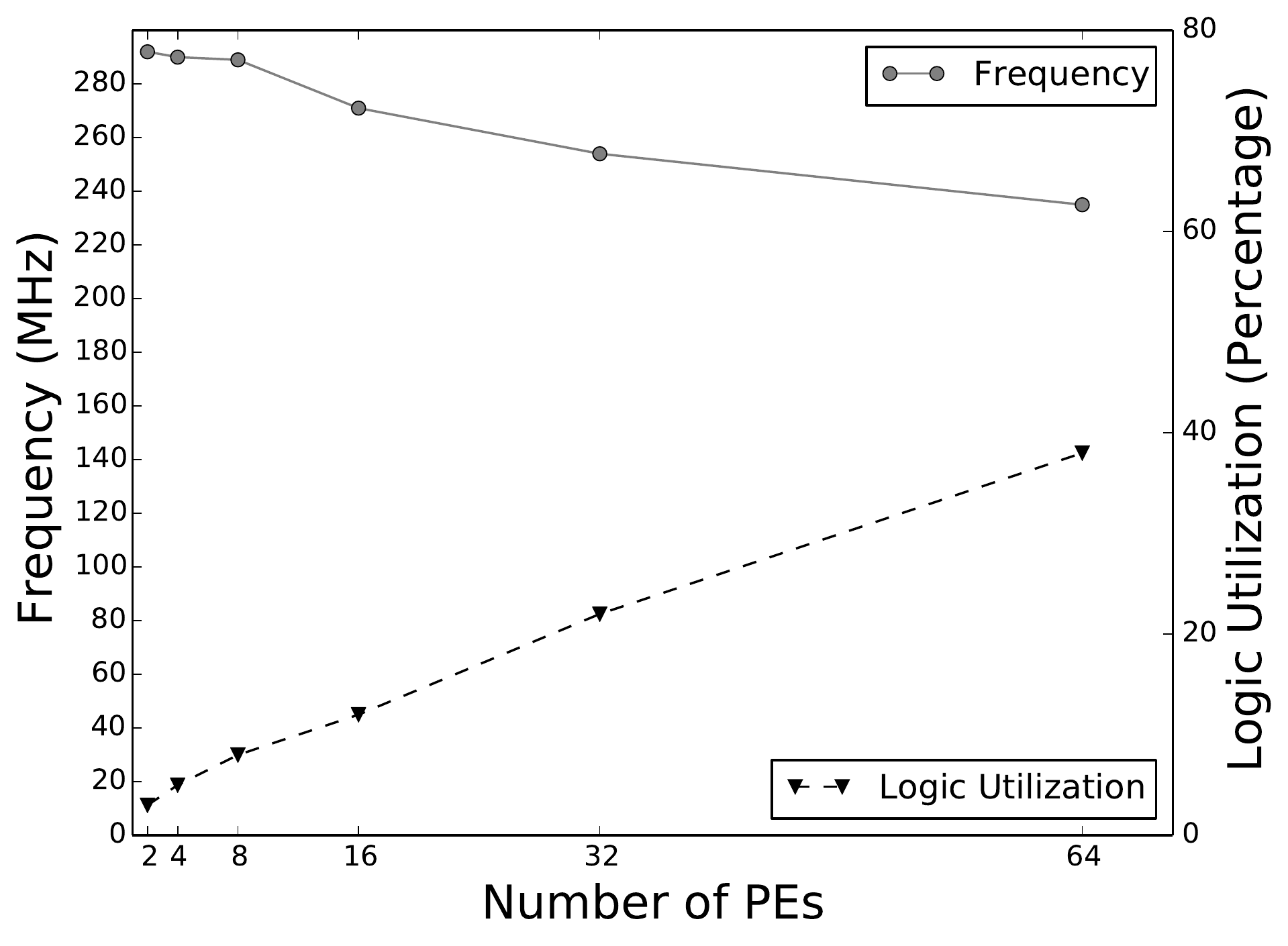}
  
\end{minipage}
\caption{\small Sensitivity Analysis. (Left) GFLOPS rate as we
  increase the number of floating point units in REAP and the
  CPU. (Right) The frequency (Mhz) and logic utilization (percentage)
  rate as we increase the number of pipelines.}
\label{fig:sensitivity}
\vspace{-2.5mm} 
\end{figure}

 \textbf{GFLOPS-Analysis.} One important
goal of our design is to improve the throughput by better utilizing
the resources (e.g. floating-point units). Figure
\ref{fig:sensitivity} shows the GFLOPS for both REAP and Intel MKL
doing SpGEMM. The GFLOPS are normalized with the number of
floating-point units available to both our design and the CPU. We only
show the median, GEOMEAN, 25th percentile and 75th percentile results
across all benchmarks.
We observe that for the same available floating-point units, REAP
achieves higher GFLOPS for all the cases. REAP better utilizes its
resources thanks to the preprocessing done by the CPU. Further, REAP's
GFLOPS scales better with the increase in the number of floating-point
units compared to the CPU.

\begin{figure}
\centering

\begin{minipage}{.25\textwidth}
  \centering
  \includegraphics[width=1\textwidth,angle=0]{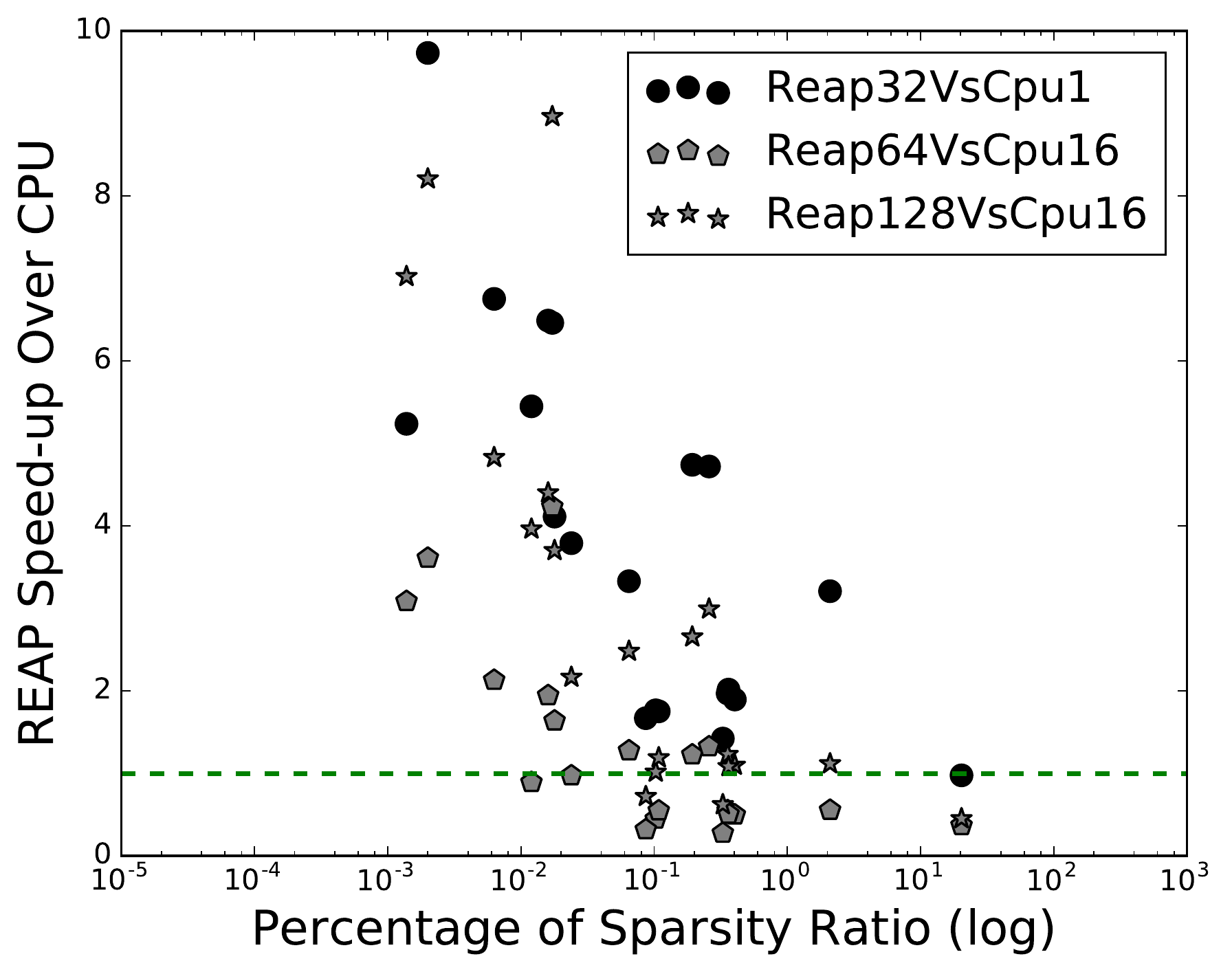}
\end{minipage}%
\begin{minipage}{.25\textwidth}
  \centering
  \includegraphics[width=1\textwidth,angle=0]{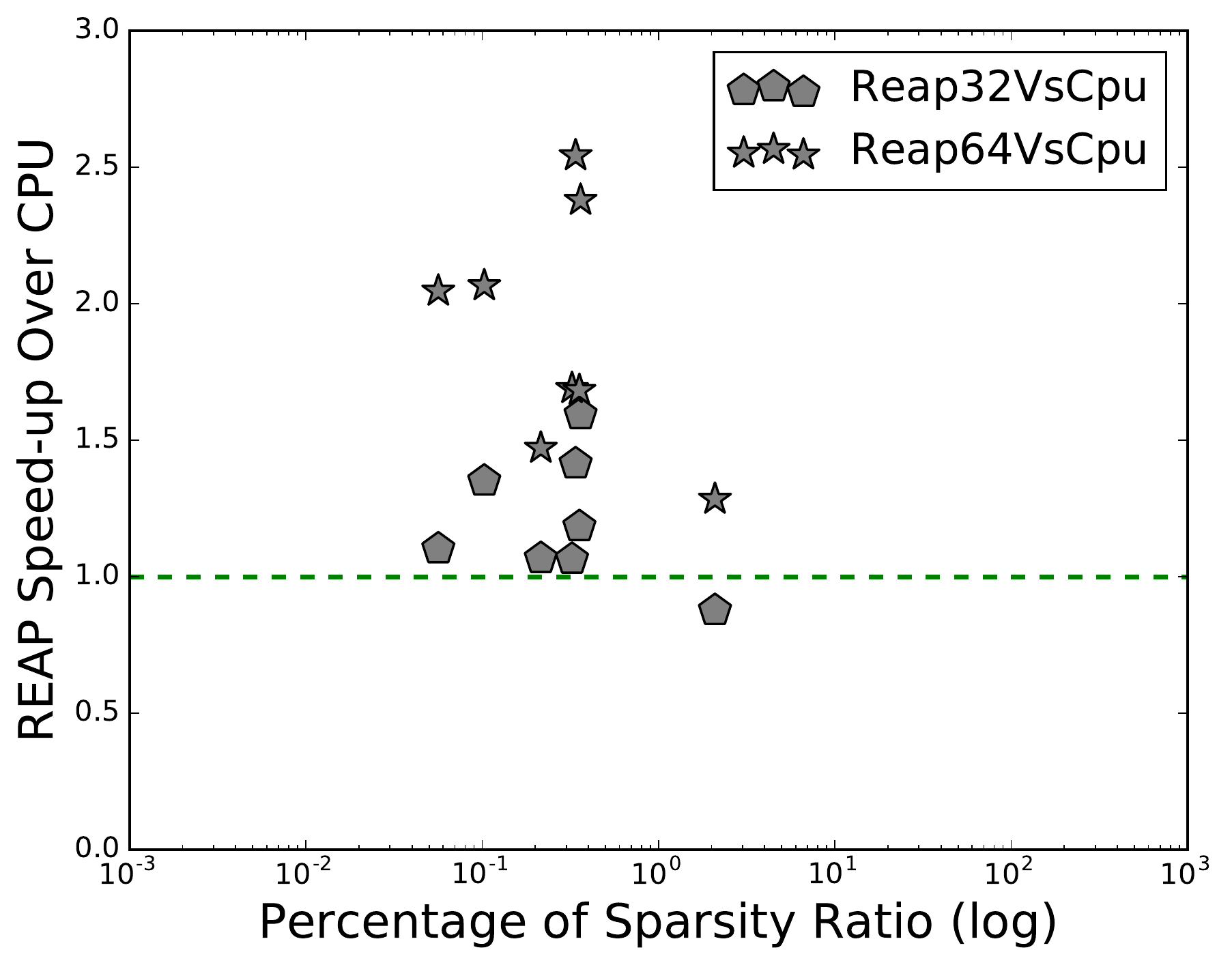}
\end{minipage}
\caption{Sensitivity to sparsity ratio for SpGEMM and sparse Cholesky.}
\label{fig:sparsity}
\end{figure}

\textbf{Hardware Scalability.} Figure \ref{fig:sensitivity} also shows
how the frequency and logic utilization varies as we add more
pipelines. While the number of pipelines changed from 2 to 128, the
logic utilization has increased only 8$\times$ and the frequency has
only dropped from 280 MHz to 220 MHz. This is because we have
extensively benefited from the DSP units and on-chip memory.
The simple interconnection network between PEs and minimal use of CAMs
are other reasons for this good scalability of our FPGA design.

\textbf{Sensitivity to sparsity.} Another interesting observation is
in Figure~\ref{fig:sparsity} where the X-axis shows the density of the
input matrix (number of non-zeros / total number of elements)*100) in
log scale and Y-axis shows the relative speedup of different
variations of REAP compared to the CPU versions. For SpGEMM, our
design achieves a relatively better speedup when the input matrix is
more sparse. In other words, REAP favors sparse matrices. The dashed
line shows where the CPU version beats the REAP. CPU beats REAP only
for the case where the matrix is relatively denser.

\subsection{Performance Evaluation with Cholesky}
\label{cholesky-res}
\begin{figure*}
\includegraphics[width=1\textwidth,angle=0]{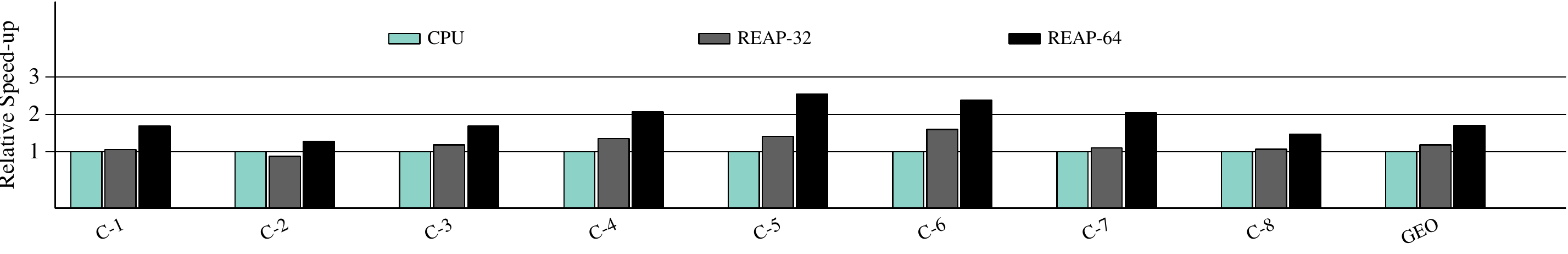}
\caption{\small The speedup of REAP designs compard to CPU execution
  of CHOLMOD on a single core for sparse Cholesky factorization.}
\label{fig:resultchol}
\end{figure*}

We compare our design of Cholesky with CHOLMOD~\cite{CHOLMOD:2008},
the state-of-the-art library for sparse Cholesky
factorization. CHOLMOD offers different optimizations and
configurations for sparse Cholesky factorization while supporting both
symbolic and numerical factorization.
The library supports a variety of options for the factorization such
as simplicial or supernodal, LL\textsuperscript{T} or
LDL\textsuperscript{T}.
For a fair comparison, we compare ourselves against simplicial,
LL\textsuperscript{T} implementation with no-ordering, which is the
closest to our implementation. Besides, both our design and CHOLMOD
benefit from the symbolic analysis. We have not included the time
spent to build the elimination tree. We compare REAP results with the
CHOLMOD configuration that runs only \textit{numeric computation}.

\begin{figure}
  \includegraphics[width=0.5\textwidth,angle=0]{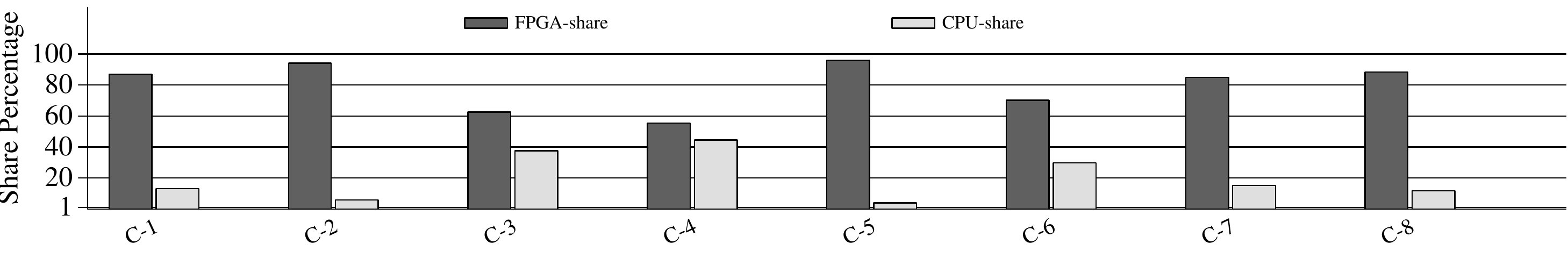}
  \caption{\small Percentage of the time spent in CPU symbolic
    analysis and preparation and in computation on the FPGA for our
    REAP-32 design for sparse Cholesky.}
  \label{fig:cholprep}
\end{figure}

\textbf{REAP variants for Cholesky.} We have two design variants.

\textbf{REAP-32}. This design has 32 pipelines and runs at 250 MHz. We
set the DRAM bandwidth to match the DRAM bandwidth for a single-core
CPU. Each PE performing the dot product is equipped with 8
multipliers. We set the bundle size and CAM size as 32.

\textbf{REAP-64} This design has 64 pipelines and runs at 238 MHz. The
DRAM bandwidth is scaled to reflect the DRAM bandwidth available for
16-cores CPU. We have also doubled the number of multipliers (\ie, 16)
for the PEs performing the dot product.  All other parameters such as
bundle size and CAM sizes are same as REAP-32 above.

Figure~\ref{fig:resultchol} shows the relative speedup of REAP
designs compared to CHOLMOD on a single core CPU.  REAP-32 design
outperforms the CPU for all but one benchmark with a geometric mean
speedup of 1.18$\times$.  REAP-64 outperforms the CPU version for all
the benchmarks on average (geometric mean) speedup of 1.85$\times$.
Unlike SpGEMM, the performance of the sparse Cholesky kernel is
limited due to the data dependencies inherently present in the algorithm.

Figure~\ref{fig:cholprep} shows the percentage of time spent in CPU
preprocessing and FPGA computation similar to our experiments with
SpGEMM. FPGA execution time significantly dominates the CPU execution
time for Cholesky. All the numeric computation take places in the FPGA
and CPU only does the symbolic computation where no floating point
operation involves.

To study the impact of dependencies on the performance of our designs,
we measured the change in performance with the increase in pipelines
and measured the idle time across all pipelines. We observed as we
increase the number of pipelines increases, the idle cycles increase
almost linearly with the number of pipelines. Hence, adding more
resources is not going to help to improve the performance of Cholesky.
There is active research in overcoming the issue of dependencies for
matrix factorization, which are orthogonal to our work.

\subsection{REAP with OpenCL HLS Designs}
To evaluate the benefits of preprocessing with REAP's RIR bundles with
an OpenCL HLS framework, we used an Intel Programmable Acceleration
Card (PAC Card) running Intel's FPGA OpenCL version 1.0. However,
shared memory between the FPGA and CPU, where the CPU's load/store instructions
as well as an OpenCL kernel's read/write statements access the same memory,
is not well supported in the current Intel OpenCL toolchain. Instead, accessor
functions are required. Thus, to evaluate REAP with HLS, we first ran the first
pass on the CPU and the FPGA did the computation on the RIR bundles
generated by the first pass. Unsurprisingly, the HLS designs are
significantly slower than the hand-coded designs.  However, the
version of REAP with HLS outperforms the HLS version without any CPU
preprocessing for all benchmarks and with a geometric mean of 16\% and
35\% for SpGEMM and Cholesky, respectively, on average across all the
benchmarks. It demonstrates that the idea of synergistic CPU-FPGA
execution is beneficial to both hand-coded designs and HLS tools.

\section{Related Work}
There exists a large body of works on optimizing the performance of sparse
linear algebra kernels~\cite{chou:tensoralgebra, Bolz:spgpu, SalitaSombatsiri2019, sparsedeeplearning, vectorcacheSpMV}. We
restrict our comparison to prior research that optimizes SpGEMM and
Cholesky.

\textbf{SpGEMM on CPUs.} Intel Math Kernel Library
(MKL)~\cite{intelmkl} provides math routines for sparse multiplication
that are widely used and highly optimized for the CPU with OpenMP and
vectorized with AVX extensions. There is a large body of work on
building cache-friendly and distributed
algorithms~\cite{Sulatycke:cputheoryspgemm:1998, sulatycke:cpuspgemm,
  Akbudak:localityspgemm, buluc:hypersparse,
  buluc:cpubandwidth,Nishtala:cacheblockspmv:2007,
  cevdet:cpulargescale:2016}, system level optimizations with
parallelism on modern
hardware~\cite{saule:cpuspgemm,matam:spgemm-modern}, and exploration
of various sparse formats for efficient
storage~\cite{Li:SMAT:pldi:2013} \cite{Buluc:dcsr,liu:sparsestorage,
  Kincaid:ellpack, Saad:csr, Kourtis:csx:2011,
  Belgin:patternbasedformat,buluc:sparseblock:spaa}.
Recent work by Zhen et al~\cite{Zhen:IA-SOGEMM:ICS:2019} proposes a
framework that accelerates the SpGEMM kernel for both CPU and GPUs by
automatically using the best sparse format and the best parallel
algorithm based on the characteristics of the input.

\textbf{SpGEMM on GPUs.} Although GPUs can outperform CPUs for dense
kernels, sparse computation introduces new challenges. Commercial
sparse libraries such as cuSPARSE~\cite{cusparse} and CUSP~\cite{cups}
apply GPU specific optimization to improve the performance of sparse
computation.  Recent research~\cite{liu:gpu:spgemmframe,
  liu:2014:gpuspgemm, matam:spgemm-modern} identifies three aspects to
improve performance on a GPU: memory allocation for the result matrix,
random memory accesses due to parallel operations, and load balancing
that has to consider sparsity.

\textbf{SpGEMM on FPGAs.} Prior work has also explored sparse
matrix-matrix multiplication architectures on
FPGAs~\cite{lin:fpgaspgemm:exploration,Nurvitadhi:fpgaanalytics:date:2016,yavits:associateprocessor:2017,yavits:spgemmcam}. They
have focused on user control on the number of PEs and the block size
to obtain a particular energy-delay product or a power-delay product.
Similar to our designs, they also exploit dedicated DSP blocks and
on-chip memory for accelerating SpGEMM computation.  However, they
assume that the FPGA on-chip resources can accommodate the entire
matrices, which is not practical for many real-world scenarios.
Recent work also uses 3D-stacked logic-in-memory to accelerate the
processing of sparse matrix data~\cite{zhu:spgemm:3Dstacked}. Use of
such logic-enhanced CAMs help with index matching, which is
orthogonal to our work.

\textbf{GEMM/SpGEMM with ASICs.} Google's tensor processing unit (TPU)
ASIC accelerates the inference phase of neural
networks~\cite{jouppi:tpu}. The building block of a TPU is a
matrix-multiply unit. However, TPU is designed for dense
matrices. Similarly, Extensor~\cite{Hegde:2019:extensor:micro} is an
ASIC that supports high-dimensional sparse data known as tensors and
helps to match the non-zero elements quickly.

\textbf{Closely related work.}
SMASH~\cite{Kanellopoulos:2019:Smash:micro} that supports a
software-hardware solution is closely related. The software encodes
the sparse data as a \textit{hierarchy of bitmaps}. This encoding is
then used by the hardware to avoid a lot of unnecessary accesses and
reduce the overhead.  Also related are recent approaches for
outer-product based matrix
multiplication~\cite{pal:outerspace,mishra:sparsefpga:dac:2017}.  Like
our approach, they also include multiply and merge phases and generate
partial products.  The use of the outer-product algorithm eliminates
the index-matching phase and allows maximum reuse of the input
matrices. However, the main drawback of an outer-product formulation
is the accumulation phase. The partial products that belong to the
entire final result matrix need to be stored. Accumulation of partial
product waits till all partial products are ready.  Hence, it
introduces complexities with the accumulation of partial results,
which makes the latency of the entire algorithm quickly dominated by
it.  In contrast, our row-by-row formulation strikes a good balance
with the reuse of data and the accumulation of partial products as
they belong to one row. Hence, our approach provides better throughput
compared to an outer-product formulation.

\textbf{Cholesky factorization.} Generating efficient sparse kernel
for Cholesky factorization an active research area in the high
performance computing community.  TACO \cite{chou:tensoralgebra} is a
code generation framework that helps generate sparse kernels in
general.  CHOLMOD \cite{CHOLMOD:2008} provides a comprehensive
framework for sparse linear solver system including high-performance
sparse Cholesky for both CPUs and GPUs. CHOLMOD provides highly
optimized sequential implementations. Intel MKL's
paradiso~\cite{schenk:paradiso} and SuperLU
\cite{Demmel:cholparallel:siam:1999} provide parallel implementations
of Cholesky for shared-memory architectures. There are recent efforts
to use inspector-executor paradigm to generate efficient sparse
kernels for Cholesky and other sparse kernels in
software~\cite{Cheshmi:2017:sympiler,Cheshmi:2018:parsy}.  Prior
research has also explored the use of GPUs to improve the performance
of Cholesky
factorization~\cite{Rennich:choleskygpu,Vuduc:limitgpuchol}.

Although there are reasonable projects for accelerating Cholesky on
CPUs, there is a lack of work on designing FPGAs for sparse Cholesky
as it is challenging to get performance with it.  An exception is a
prior work that analytically shows that FPGA have the potential to
improve the performance of sparse Cholesky~\cite{Yichun:fpgachol}.  In
contrast, this paper shows a generic design primitive that accelerates
Cholesky and related sparse kernels with some preprocessing performed
by the CPU.

\section{Conclusion}

We introduce REAP, a system for high performance and efficient sparse
linear algebra. REAP uses a co-operative strategy between a CPU and
FPGA, where the CPU re-organizes the sparse matrices from a standard
format to one that can be streamed into the FPGA. The FPGA then
performs the computations in a systolic-style, taking advantage of the
inherent parallelism of matrix operations.  Our evaluation over a
diverse set of matrices demonstrates our co-operative approach
significantly improves the performance of SpGEMM and Sparse Cholesky
factorization compared to state-of-the-art CPU implementations, even
when the FPGA and a multi-core CPU have an equal number of floating
point units.  A key take away from our research is that increasing
bandwidth alone does not obtain the highest performance; having the
CPU reformat the data into a more sequential access pattern gives the
FPGA a huge performance boost that greatly outweighs the formatting
cost.  We hope our co-operative techniques inspire the HPC community
to build applications in this style of design to better exploit HBM as
it becomes more pervasive and affordable.

\section*{Acknowledgment}

This paper is based on work supported in part by NSF CAREER Award CCF–1453086, NSF Award CCF-1917897, NSF Award CCF-1908798, NSF CNS-1836901,  NSF OAC-1925482, and Intel Corporation.

\balance
\renewcommand\refname{References}

\bibliographystyle{IEEEtran}
\bibliography{fpga}

\end{document}